\documentclass[letterpaper,twocolumn,10pt]{article}
\usepackage{usenix2019}
\usepackage{graphicx, amsfonts}
\usepackage{color, multirow, array, footnote, subfigure, bigints}
\usepackage{paralist} 
\usepackage{enumitem}
\usepackage[symbol]{footmisc}
\usepackage{filecontents}
\usepackage{amsfonts} 
\usepackage{filecontents}

\begin{document}

\title{Elastic RAID: When RAID Meets SSDs with Built-in Transparent Compression}

\author{Zheng Gu$^\dagger$, Jiangpeng Li$^\dagger$, Yong Peng$^\dagger$, Yang Liu$^\dagger$, and Tong Zhang$^\dagger$$^\ddagger$\\[6pt] $^\dagger$ {\large ScaleFlux Inc., CA, USA}\\[3pt] $^\ddagger$ {\large Rensselaer Polytechnic Institute, NY, USA}}
\maketitle

\begin{abstract}
This paper studies how RAID (redundant array of independent disks) could take full advantage of modern SSDs~(solid-state drives) with built-in transparent compression. In current practice, RAID users are forced to choose a specific RAID level (e.g., RAID~10 or RAID~5) with a fixed storage cost vs.~speed performance trade-off. Commercial market is witnessing the emergence of a new family of SSDs that can internally perform hardware-based lossless compression on each 4KB LBA~(logical block address) block, transparent to host OS and user applications. Beyond straightforwardly reducing the RAID storage cost, such modern SSDs make it possible to relieve RAID users from being locked into a fixed storage cost vs.~speed performance trade-off. The key idea is simple: RAID systems opportunistically leverage higher-than-expected runtime user data compressibility to enable dynamic RAID level conversion to improve the speed performance without compromising the effective storage capacity. This paper presents design techniques to enable and optimize the practical implementation of such {\it elastic} RAID systems. For the purpose of demonstration, we implemented a Linux software-based elastic RAID prototype that supports dynamic conversion between RAID~5 and RAID~10. Compared with a baseline software-based RAID~5, under sufficient runtime data compressibility that enables the conversion from RAID~5 to RAID~10 over 60\% user data, the elastic RAID could improve the 4KB random write IOPS (IO per second) by 42\% and 4KB random read IOPS in degraded mode by 46\%, while maintaining the same effective storage capacity. 
\end{abstract}

\section{Introduction}
\label{sec:introduction}
This paper studies the implementation of RAID~(redundant array of independent disks)~\cite{patterson1988case, chen1994raid} over SSDs~(solid-state drives) with built-in transparent data compression. As one of the best known computing system design techniques, RAID plays an important role in building reliable computing infrastructure. In current practice, when deploying a RAID system, users must choose~(and subsequently stick with) one specific RAID level after, often painfully, deliberating the trade-off between the storage cost and speed performance. For example, between RAID~10 and RAID~5, the two most popular RAID levels, RAID~10 achieves a higher I/O speed performance, in terms of IOPS (I/O requests per second) and average/tail latency, at the penalty of a higher data storage cost, while RAID~5 reduces the data storage cost by sacrificing the I/O speed performance. Such a storage cost vs.~speed performance trade-off is inherent in the design of RAID, regardless whether its implementation is software-based~(e.g., Linux mdraid~\cite{Linux-RAID-link} and Btrfs RAID~\cite{rodeh2013btrfs}) or hardware-based~(e.g., RAID controller card~\cite{Broadcom-RAID-link}).

Commercial market currently witnesses the rise of SSDs with the built-in transparent data compression capability~\cite{ScaleFlux-link, Seagate-FMS-19}. Such modern SSDs internally carry out hardware-based compression on each 4KB LBA~(logical block address) block, and could expose a logical storage space that is (much) larger than their internal physical NAND flash memory storage capacity. Evidently, one could deploy a RAID system~(regardless its RAID level) over such SSDs to reduce the effective storage cost without any changes of the RAID implementation and any degradation of the RAID speed performance. This paper shows that, beyond straightforwardly reducing the storage cost, SSDs with built-in transparent compression brings an unique opportunity to improve the RAID speed performance by elastically mixing different RAID levels~(e.g., RAID~5 and RAID~10) in adaptation to the runtime user data compressibility variations. Its basic idea is simple: Suppose we deploy a RAID 5 over multiple SSDs with the total physical storage capacity of 32TB and format the RAID logical storage capacity as 64TB, i.e., we expect that the average data compressibility is about 2:1 and hence aim at leveraging the intra-SSD transparent compression capability to reduce the effective RAID storage cost by up to 2$\times$. Of course, the storage system must accordingly adjust its monitoring and management to prevent out-of-space failure under worse-than-expected data compressibility. If the runtime data compressibility exceeds 2:1, the RAID system could {\it opportunistically} convert the protection of some user data from RAID~5 to RAID~10 in order to improve the RAID speed performance, while still maintaining the total 64TB effective RAID data storage capacity. As the runtime data compressibility dynamically varies, the RAID system adaptively adjusts the mixture of RAID~5 and RAID~10. Such an {\it elastic RAID} design strategy opportunistically utilizes the runtime {\it residual} data compressibility to improve the speed performance without compromising the effective storage capacity. If users demand double drive failure protection, elastic RAID will dynamically mix RAID~6 and triple replication. For the purpose of simplicity, this paper focuses on the case of elastic mixture of RAID~5 and RAID~10 in one RAID system, and the proposed design solutions could be readily extended to the case of elastic mixture of RAID~6 and triple replication.

In spite of its simple concept, the practical implementation of elastic RAID is subject to two major issues: (i) Implementing RAID level conversion: RAID~5 and RAID~10 have different data mapping and occupy different amount of storage capacity, which makes it nontrivial to dynamically convert between RAID~5 and RAID~10 on the same array of SSDs. Meanwhile, RAID level conversion incurs data copy/move/delete operations, and it is highly desirable to minimize the conversion-induced operational overhead and its impact on foreground user read/write requests. Moreover, we must retain the drive failure protection during the RAID level conversion. This paper presents a {\it bloated stripe allocation} design approach to facilitate the implementation of dynamic RAID level conversion. It minimizes the RAID level convention operational overhead and meanwhile enables fine-grained conversion on the per-stripe basis, i.e., RAID level conversion can be carried out on each individual RAID stripe independently from all the other stripes. Such per-stripe RAID level conversion enables the most graceful adaptation to the runtime data compressibility variation. 
(ii) Scheduling RAID level conversion: The second issue is how to dynamically schedule RAID level conversion to maximize the speed performance gain by keeping as many hot data as possible under the RAID~10 protection. This paper presents a strategy to realize RAID level conversion both {\it proactively} in adaptation to workload characteristics variation and {\it reactively} in response to data compressibility change. This scheduling strategy takes per-stripe data compressibility into account to increase the coverage of RAID~10, and leverages the principle of the well-known second chance algorithm~\cite{silberschatz2013operating} to reduce the implementation overhead. In addition to addressing the above two major issues, this paper presents a simple technique to further improve the implementation efficiency of elastic RAID. It is motivated by the observation that RAID parity tends to have much worse compressibility than user data, because parity calculation could largely destroy the data entropy. This suggests that a RAID~5 stripe does not necessarily always consume less physical storage space~(post in-storage transparent compression) than its RAID~10 counterpart. Accordingly, an elastic RAID system could proactively convert a RAID~5 stripe to its RAID~10 counterpart if the conversion does not increase the physical storage space usage. 

For the purpose of demonstration, we implemented a Linux software elastic RAID prototype in support of the mixture of RAID~5 and RAID~10. This prototype was developed by modifying/enhancing the existing Linux mdraid~\cite{Linux-RAID-link} to incorporate the proposed design techniques and meanwhile enhance the support of multi-threaded operations. We carried out experiments by deploying the software elastic RAID over commercial SSDs with built-in transparent compression from Scaleflux Inc.~\cite{ScaleFlux-link}. We applied the widely used FIO~(flexible I/O tester) tool~\cite{FIO-link} to generate heavy I/O workloads and collect the IOPS and tail latency results. When operating in the RADI~5 only mode, our elastic RAID implementation could noticeably outperform the state-of-the-art software RAID~5 product RAIDIX~\cite{RAIDIX-link}, and both RAIDIX and our elastic RAID achieve $\sim10\times$ higher IOPS than the Linux mdraid. We further carried out experiments to evaluate the effect of elastic RAID~5 and RAID~10 mixture, and the results well demonstrate its efficacy on improving the RAID speed performance without compromising the effective RAID storage capacity. For example, compared with the baseline that operates in the RAID~5 only mode, converting 20\% and 60\% user data from RAID~5 to RAID~10 will improve the 4KB random write IOPS by 10\% and 42\%, respectively. When the RAID system operates in the degraded mode (i.e., one SSD is off-line), converting 20\% and 60\% RAID~10 user data from RAID~5 to RAID~10 will improve the 4KB random read IOPS by 12\% and 46\%, respectively. The experimental results show that a small increase of user data compression ratio could enable a significant increase of RAID~10 coverage. For example, the RAID~10 coverage could improve by over 40\% if user data compression ratio slightly increases from 1.2 to 1.4. The experimental results also show that the RAID level conversion can be carried out with a very high throughput and its impact on the RAID system speed performance is very small~(e.g., less than 5\%) even under very heavy foreground user I/O workloads. This work well demonstrates that the emerging SSDs with built-in transparent compression for the first time make it practically feasible for a RAID system to opportunistically mix different RAID levels to improve the speed performance without compromising the effective data storage capacity.

\section{In-Storage Transparent Compression}
\label{sec:SSD}
\begin{figure}[t]
  \centering
  \includegraphics[width = \linewidth]{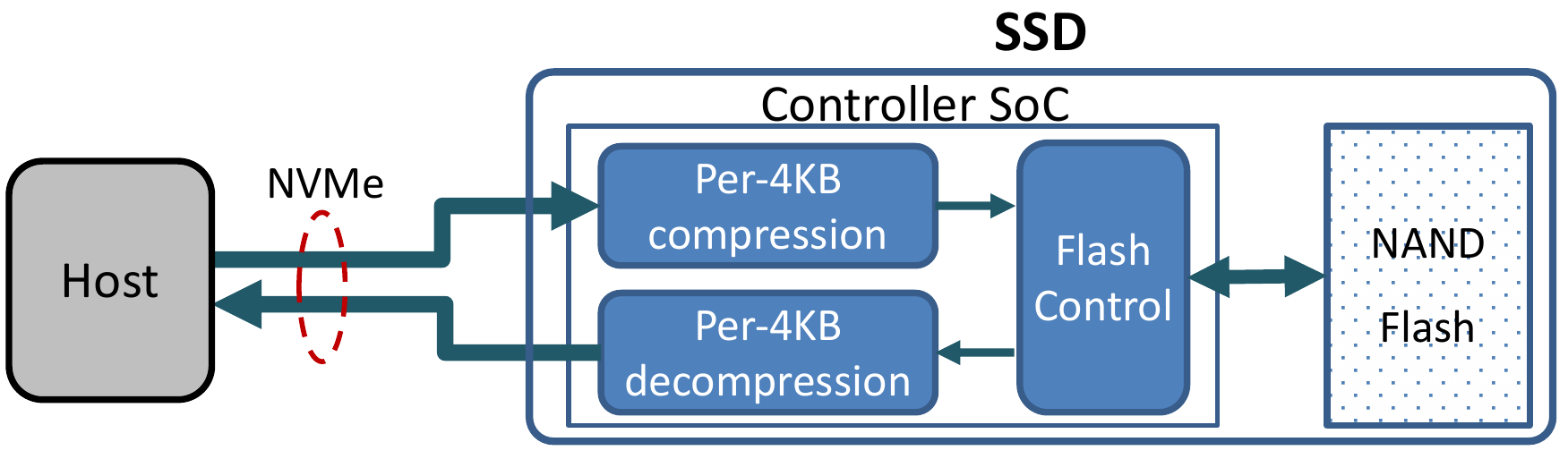}
  \caption{Illustration of an SSD with built-in transparent compression.}
  \label{fig:SSD}
\end{figure}

Fig.~\ref{fig:SSD} illustrates an SSD with built-in transparent compression: Its controller SoC~(system on chip) performs  (de)compression on each 4KB LBA data block along the I/O path, and manages the placement of all the post-compression variable-length data blocks on the NAND flash memory. The in-storage per-4KB data compression is transparent to the host that accesses the SSD as a normal block data storage device through a standard I/O interface~(e.g., NVMe or SATA). The per-4KB (de)compression latency of the hardware engine inside the SSD controller SoC can be well below 5$\mu$s, which is over 10$\times$ shorter than the TLC/QLC NAND flash memory read latency~($\sim$50$\mu$s and above) and write latency~($\sim$1ms and above). Meanwhile, the hardware (de)compression engines could easily achieve a throughput higher than the aggregated bandwidth of back-end NAND flash memory chips~(e.g., 4$\sim$8GB/s). Therefore, SSDs with built-in transparent compression can maintain the same IOPS and latency performance as traditional SSDs without built-in compression capability. In fact, by reducing the write stress on NAND flash memory through compression, such SSDs could have~(much) lower GC~(garbage collection) overhead, leading to (much) higher IOPS under write-intensive workloads. For example, under heavy 4KB random writes with 100\% LBA span, traditional NVMe SSDs could achieve up to 200K$\sim$300K IOPS, while NVMe SSDs with built-in transparent compression (e.g., the one from ScaleFlux Inc.~\cite{ScaleFlux-link}) could achieve over 600K IOPS under 2:1 user data compression ratio.

\begin{figure}[hbtp]
  \centering
  \includegraphics[width = 0.9\linewidth]{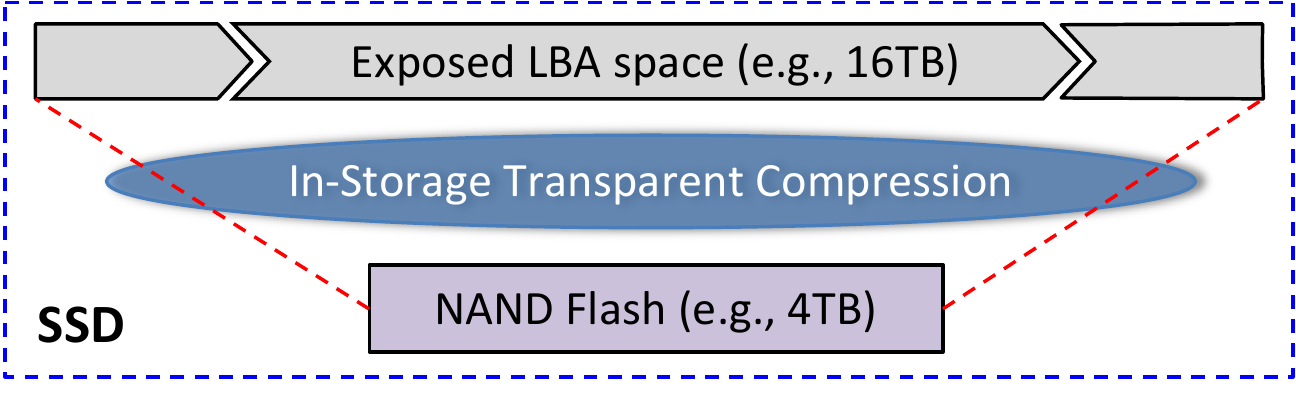}
  \caption{Illustration of the expanded LBA space exposed by SSDs with built-in transparent compression.}
  \label{fig:ExposedLBA}
\end{figure}

To allow the host materialize the benefit of in-storage transparent data compression, such modern SSDs could expose an expanded LBA logical storage space that is larger (e.g., by 2$\times$ or 4$\times$) than its internal physical NAND flash memory storage capacity, as illustrated in Fig.~\ref{fig:ExposedLBA}. Given the runtime data compressibility variation, such SSDs with expanded LBA space may possibly run out of physical storage space before their exposed logical storage space has been used up by the host. Hence, to avoid running into the {\it out-of-space} error, the host must keep monitoring the SSD physical storage space usage and accordingly make its storage management aware of the runtime physical storage space usage, just like when using any thin-provisioned storage systems. 

\section{Proposed Elastic RAID}
\label{sec:elasticRAID}
When users deploy RAID over SSDs with built-in transparent compression, they should format its logical storage capacity based on the expected/estimated data compressibility in order to reduce the effective data storage cost. Let us consider RAID~5 over $n+1$ SSDs, and let $C_{flash}$ denote the NAND flash memory capacity of each SSD~(excluding its internal over-provisioned storage capacity reserved for GC). Throughout this paper, we define the compression ratio as the pre-compressed data size being divided by the post-compressed data size (i.e., a larger compression ratio corresponds to a higher data compressibility). Let $\alpha_{exp}\ge 1$ denote the expected average data compression ratio over the entire RAID~5~(including both user data and RAID~5 parity). Since all the user data and RAID~5 parity are evenly striped across the $n+1$ SSDs, all the SSDs will experience the same average data compression ratio. Hence, the logical storage capacity of the RAID~5 system should be formatted as $C_{RAID}=\alpha_{exp}\cdot n\cdot C_{flash}$, i.e., users expect to increase the effective RAID storage capacity by up to $\alpha_{exp}\times$ via deploying SSDs with built-in transparent compression. In practice, users tend to purposely underestimate the value of $\alpha_{exp}$ in order to better embrace unexpected data compressibility degradation and simplify the storage system management. Let $\alpha$ denote the runtime average compression ratio of all the data (both user data and parity) on RAID~5. Once the runtime data compressibility is better than expected~(i.e., $\alpha>\alpha_{exp}$), it will leave a certain amount of physical NAND flash memory storage space unused. Intuitively, we could utilize such {\it opportunistically available} physical data storage capacity to convert RAID~5 to RAID~10 over a portion~(or even all) of user data. This can contribute to improving the RAID I/O speed performance including IOPS and latency, while still maintaining the same RAID logical storage capacity of $C_{RAID}$. 

Fig.~\ref{fig:RAID} further illustrates the basic concept of such elastic RAID: 
\begin{figure}[b]
  \centering
  \includegraphics[width = 0.9\linewidth]{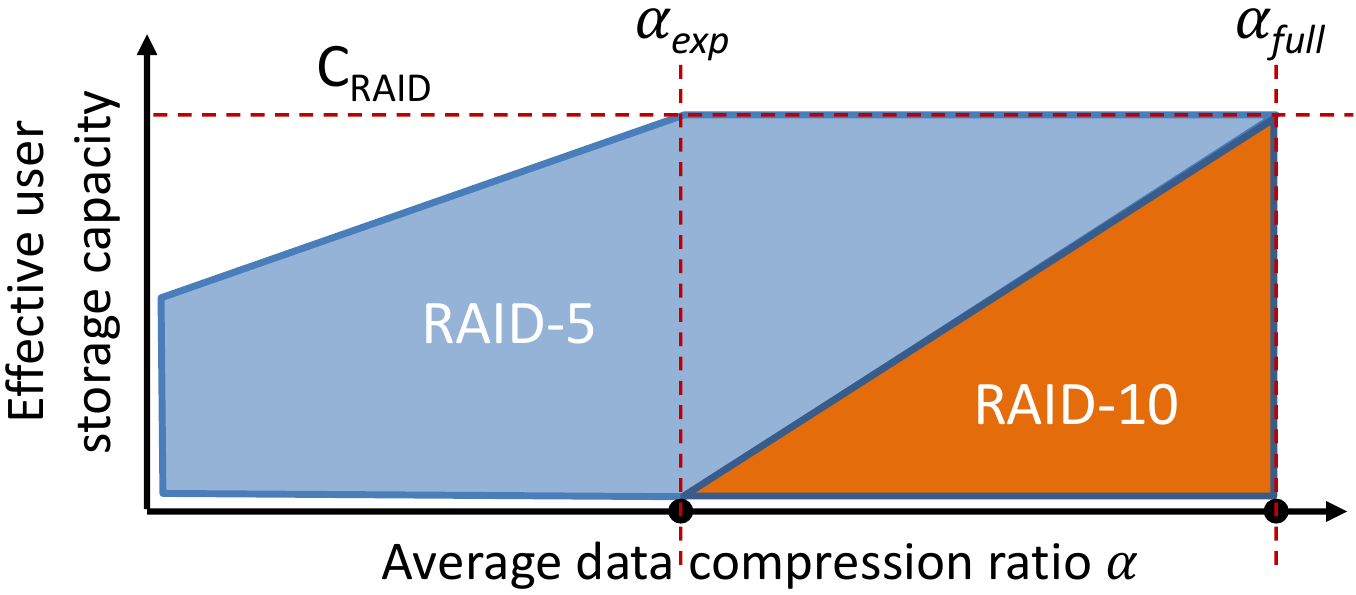}
  \caption{Illustration of elastic RAID enabled by in-storage transparent compression.}
  \label{fig:RAID}
\end{figure}
To materialize the storage cost reduction enabled by deploying SSDs with built-in transparent compression, an elastic RAID system exposes a logical storage capacity of $C_{RAID}$ to the host. 
When the runtime average data compression ratio $\alpha$ is less than $\alpha_{exp}$, the elastic RAID system entirely operates as a classical RAID~5 to transform 100\% data compressibility into the storage cost reduction, as shown in Fig.~\ref{fig:RAID}. Being proportional to the average data compression ratio $\alpha$, the effective data storage capacity that is truly consumable by the users is lower than the logical storage capacity $C_{RAID}$ exposed by the elastic RAID. Hence, users must closely monitor the true physical storage capacity usage of SSDs in order to avoid out-of-space errors. Once the average runtime data compression ratio $\alpha$ reaches $\alpha_{exp}$, the effective data storage capacity will reach $C_{RAID}$~(i.e., all the logical storage space exposed by the RAID system could be truly consumed by the users). As $\alpha$ exceeds $\alpha_{exp}$, the effective data storage capacity will remain as $C_{RAID}$ and the residual data compressibility~(i.e., $\alpha-\alpha_{exp}$) will be exploited to enable RAID~5 to RAID~10 conversion, as illustrated in Fig.~\ref{fig:RAID}, where $\alpha_{full}$ denotes the data compression ratio under which all the user data could be protected by RAID~10. 

In summary, as the runtime average data compression ratio $\alpha$ dynamically varies between [1, $\alpha_{exp}$], the effective data storage capacity of elastic RAID will proportionately change and all the user data are protected by RAID~5; as $\alpha$ dynamically varies between [$\alpha_{exp}$, $\alpha_{full}$], elastic RAID will accordingly adjust the mixture of RAID~5 and RAID~10 while its effective data storage capacity remains as $C_{RAID}$; once $\alpha$ exceeds $\alpha_{full}$, all the user data are protected by RAID~10 and the effective data storage capacity remains as $C_{RAID}$. Although elastic RAID is built upon a simple idea, its practical and efficient implementation poses unique challenges. The remainder of this section will present techniques to address the major implementation challenges on the realization and scheduling of RAID level conversion, and one additional technique to further improve its implementation efficiency.   

\subsection{Realization of RAID Level Conversion}
\label{sec:RAIDimplementation}
Compared with conventional RAID implementation, one unique challenge of elastic RAID is the realization and management of dynamic conversion between different RAID levels. Let $n+1$ denote the total number of SSDs in the RAID system, and $\mathcal{P}_i$ denote the LBA space exposed by the $i$-th SSD. Let $\mathcal{L}_{usr}$ denote the LBA space exposed by the RAID system to the user. The RAID system applies a mapping $f:\ \mathcal{L}_{usr}\to \{\mathcal{P}_1,\cdots, \mathcal{P}_{n+1}\}$ to manage the user data storage over the SSD array. With different amounts of data redundancy, different RAID levels must use different mapping function $f$, leading to different user data placement over the $n+1$ SSDs. As a result, dynamic conversion between different RAID levels demands runtime varying the mapping function $f$ (and hence the placement of user data and RAID parity). This could significantly complicate the data placement and management. Meanwhile, changing the mapping function will incur SSD operational overheads in terms of data copy/move/delete operations, leading to interference with foreground user I/O requests. Elastic RAID should reduce such operational overheads as much as possible to maximize its speed performance gain. 

\begin{figure}[htbp]
  \centering
  \includegraphics[width = \linewidth]{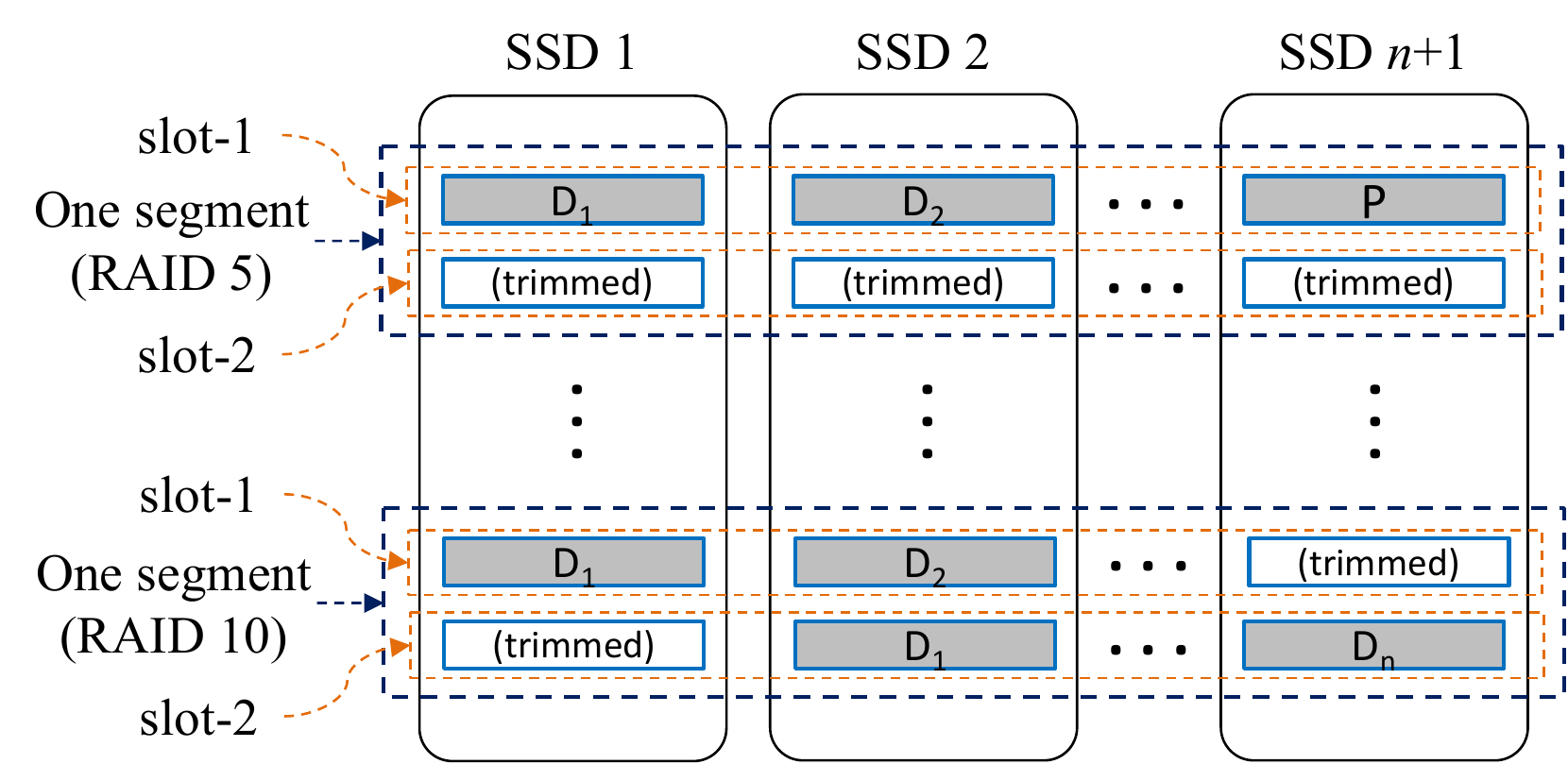}
  \caption{Illustration of the bloated stripe allocation to facilitate the conversion between RAID~5 and RAID~10.}
  \label{fig:RAIDzone}
\end{figure}

This work proposes a design technique to simplify the switching between different data mapping functions during the RAID level conversion. As pointed out above in Section~\ref{sec:SSD}, SSDs with built-in transparent compression could expose an expanded LBA space that is much larger than the physical NAND flash memory storage capacity. This enables the storage systems purposely {\it under-utilize} the SSD LBA space while still fully utilizing the physical NAND flash memory storage capacity. Leveraging this simple fact, the key idea of the proposed design technique is to trade the SSD LBA space utilization efficiency for simpler data placement and management in support of dynamic RAID level conversion. Fig.~\ref{fig:RAIDzone} illustrates this design technique in the context of elastic RAID with RAID~5 and RAID~10. Given the total $n+1$ SSDs, each RAID~5 stripe contains $n$ user data strips~(denoted as $D_1,\cdots, D_{n}$) and one parity strip~(denoted as $P$). We partition the entire LBA space of all the $n+1$ SSDs $\{\mathcal{P}_1,\cdots, \mathcal{P}_{n+1}\}$ into a large number of {\it segments}, where each segment is $2\times$ larger than one RAID~5 stripe. As illustrated in Fig.~\ref{fig:RAIDzone}, each segment is further partitioned into two equal-sized slots denoted as {\it slot-1} and {\it slot-2}. The two slots in each segment hold different content when the corresponding data stripe is protected by RAID~5 or RAID~10:
\begin{itemize}
\item In case of RAID~5, {\it slot-1} stores the entire RAID~5 stripe and {\it slot-2} is empty~(hence all the LBA data blocks in {\it slot-2} are trimmed). The utilization of the expanded SSD LBA space is 50\%.
\item In case of RAID~10, {\it slot-1} and {\it slot-2} each stores one copy of stripe user data $D_1,\cdots, D_{n}$. In each slot, one strip is left unused and hence can be trimmed. The utilization of the expanded SSD LBA space is $n/(n+1)$. 
\end{itemize}

By under-utilizing the SSD LBA space through the bloated stripe allocation, we can easily switch the data mapping function in support of dynamic RAID level conversion. Accordingly, the RAID system mapping function $f$ can be decomposed into a large number of independent segment mapping functions $f^{(k)}:\ \mathcal{L}_{usr}^{(k)}\to \{\mathcal{P}_1^{(k)},\cdots, \mathcal{P}_{n+1}^{(k)}\}$, where $\mathcal{L}_{usr}^{(k)}$ denotes the user data covered by one RAID stripe and $\mathcal{P}_i^{(k)}$'s correspond to the LBA space occupied by one segment. This conveniently enables the fine-grained per-stripe RAID level conversion, i.e., RAID level conversion can be done for each individual stripe independently from all the other stripes. Within each segment, elastic RAID systems could easily realize dynamic conversion between RAID~5 and RAID~10: 
\begin{itemize}
\item To convert one segment from RAID~5 to RAID~10, we first copy all the data strips $D_i$'s from {\it slot-1} to {\it slot-2} in a skewed pattern so that two copies of the same data will not reside on the same SSD, and then trim the unused LBAs in {\it slot-1}.
\item To convert one segment from RAID~10 to RAID~5, we first calculate the parity strip $P$ based on the user data strips, write $P$ to {\it slot-1}, and finally trim all the LBA data blocks in {\it slot-2}.
\end{itemize}

During the conversion between RAID~5 and RAID~10, we never in-place update any data blocks, which can ensure the atomicity of the conversion operation and hence keep data being protected during the conversion. Moreover, such fine-grained per-stripe RAID level conversion allows the most graceful adaptation to the runtime variations of data compressibility and/or data access locality. Accordingly, the elastic RAID system should maintain a segment bitmap in which each bit records whether its associated segment uses RAID~5 or RAID~10. Moreover, since this design approach under-utilizes the SSD LBA space by up to 50\%, given the target data compression ratio $\alpha_{exp}$, each SSD should be formatted with an LBA space expansion factor of $2\alpha_{exp}$.

\subsection{Scheduling of RAID Level Conversion}
\label{sec:scheduling}
Elastic RAID must incorporate a scheduler that is responsible for two tasks: (i) determine when RAID level conversion should be triggered, and (ii) choose the stripes to which RAID level conversion should be applied. In the following, we will discuss the possible design approaches for accomplishing these two tasks.

\subsubsection{Triggering RAID Level Conversion}
\label{sec:task1}
RAID level conversion could be triggered {\it reactively} in response to data compressibility variation or {\it proactively} in adaptation to workload I/O data access characteristics variation. To maximize the speed performance gain, elastic RAID should prioritize on keeping write/read-hot data in the RAID~10 region. Moreover, when workload I/O data access locality changes over the time, RAID~10 region may no longer well cover the currently hot data, which should trigger RAID level conversion to accordingly adjust the RAID~10 region. 

We first discuss the scenario of reactively triggering RAID level conversion in response to data compressibility variation. 
The above discussion, as illustrated in Fig.~\ref{fig:RAID}, suggests that all the user data should be protected by RAID~5 until the average data compression ratio $\alpha$ reaches $\alpha_{exp}$ under which the effective RAID storage capacity will reach $C_{RAID}$. This implicitly assumes that the users always utilize 100\% of the available data storage capacity. Nevertheless, it is not uncommon that, in real-world production environment, storage capacity utilization can be well below 100\% (e.g., 80\%$\sim$90\% and even lower) due to factors such as runtime storage usage fluctuation and storage capacity over-provisioning for operational safety margin. Therefore, instead of triggering RAID level conversion only when $\alpha\in(\alpha_{exp}, \alpha_{full})$, we could opportunistically trigger RAID level conversion even under $\alpha<\alpha_{exp}$, in the case of users under-utilizing the available storage capacity. This can be realized by scheduling RAID level conversion directly based on the runtime physical storage capacity usage, other than the runtime data compression ratio. One simple scheduling approach is described as follows: The scheduler periodically samples the runtime physical storage capacity usage (denoted as $C_{util}$) in each SSD. Recall that $C_{flash}$ denotes the total physical storage capacity of each SSD, and let $C_u$ and $C_l$ denote two pre-defined parameters~(where $C_{flash}>C_u>C_l$). If $C_{util}$ does not exhibit significant variation and remains below $C_l$ in all the SSDs, then the scheduler will declare that the available storage capacity is being under-utilized, and hence will trigger the conversion to promote some regions from RAID~5 to RAID~10 to trade the under-utilized storage capacity for higher speed performance. If $C_{util}>C_u$ in any SSD and elastic RAID currently contains RAID~10 regions, then the scheduler will trigger the conversion to demote some regions from RAID~10 to RAID~5 to restore sufficient operational safety margin for sudden storage capacity usage spikes.   

To proactively trigger RAID level conversion in adaptation to data access locality variation, elastic RAID must keep monitoring the relative {\it hotness} of the data being currently covered by RAID~10. Let $\mathcal{S}$ denote the set consisting of all the stripes, and $\mathcal{S}^{(10)}$ denote the set consisting of all the RAID~10 stripes. Upon each I/O request, elastic RAID checks its bitmap to identify whether the request hits a RAID~5 or RAID~10 region. Therefore, elastic RAID could readily obtain the current RAID~10 data access hit rate $\beta$~(i.e., the probability of a data access request hitting the RAID~10 region). Given a pre-defined factor $h>1$, we claim that the RAID~10 region no longer well covers the currently hot data if
\begin{equation}
    \beta < h\cdot \frac{\big|\mathcal{S}^{(10)}\big|}{|\mathcal{S}|},
\end{equation}
where $|\cdot|$ denotes the number of elements in a set. This will accordingly trigger the RAID level conversion that aims to better align the RAID~10 region with currently hot data. 

\subsubsection{Stripe Selection for RAID Level Conversion}
\label{sec:task2}
Once the scheduler has triggered the RAID level conversion, the next task is to choose the data stripes for RAID level conversion, with the objective of keeping/converting hot stripes into the RAID~10 region. Hence, the scheduler must incorporate a hot stripe classification mechanism. To reduce the scheduling-induced implementation overhead, one could apply low-cost algorithms (e.g., the well-known {\it second chance} algorithm~\cite{silberschatz2013operating}) for hot stripe classification. One possible implementation is described below: Let ${\bf V}^{(5)}$ and ${\bf V}^{(10)}$ denote bitmaps in which one bit corresponds to one stripe in RAID~5 and RAID~10, respectively. Bitmap ${\bf V}^{(5)}$ is used to identify the candidate stripes for being promoted from RAID~5 to RAID~10, and bitmap ${\bf V}^{(10)}$ is used to identify the candidate stripes for being demoted from RAID~10 to RAID~5. The operations on these two bitmaps are summarized as follows:
\begin{itemize}
    \item Operations on ${\bf V}^{(5)}$: All the bits in ${\bf V}^{(5)}$ are periodically reset to zero. Whenever a stripe is accessed by an I/O request, its corresponding bit in ${\bf V}^{(5)}$ will be set to 1. To select a candidate stripe for promotion to RAID~10, the scheduler chooses a stripe whose bit is 1 in ${\bf V}^{(5)}$.
    \item Operations on ${\bf V}^{(10)}$: Whenever a stripe in RAID~10 is accessed by an I/O request, its corresponding bit in ${\bf V}^{(10)}$ will be set to 1. The scheduler uses the  second chance algorithm to select a candidate stripe for demotion to RAID~5, i.e., the scheduler scans ${\bf V}^{(10)}$ to find the first zero and marks the corresponding stripe as a candidate, and when it encounters a one during the scan, it will reset it to zero and move to the next bit.
\end{itemize}

If SSDs support users to query data compression ratio on the per-LBA basis, once the scheduler has chosen a candidate stripe for promotion to RAID~10 or demotion to RAID~5, it will decide whether to indeed promote/demote the chosen candidate in a probabilistic manner by incorporating the per-stripe compressibility information. This is because keeping more compressible stripes in RAID~10 helps to increase the coverage of RAID~10 and hence benefit the RAID speed performance. Since the FTL~(flash translation layer) inside SSDs must keep track the length of each compressed LBA data block, it may not be too difficult for SSDs to support such per-LBA compression ratio query. In this case, given a chosen candidate stripe for promotion to RAID~10~(or demotion to RAID~5), the scheduler accepts this candidate for promotion~(or demotion) with a probability being proportional~(or inversely proportional) to its compression ratio~(i.e., a more compressible candidate has a higher chance to be promoted into the RAID~10 region, and a less compressible candidate has a higher chance to be demoted to the RAID~5 region). Such a simple probabilistic approach could help to increase the number of hot stripes covered by RADI~10, without incurring significant CPU/memory usage during the decision process. If SSDs do not support per-LBA compression ratio query (e.g., they only support the query of the average compression ratio of the entire SSD), then the scheduler will treat all the chosen candidate stripes equally and randomly pick some or all of the chosen candidate stripes for promotion or demotion.

\subsection{Autonomous RAID Level Conversion}
\label{sec:optimization}
\begin{figure}[hbtp]
  \centering
  \includegraphics[width = \linewidth]{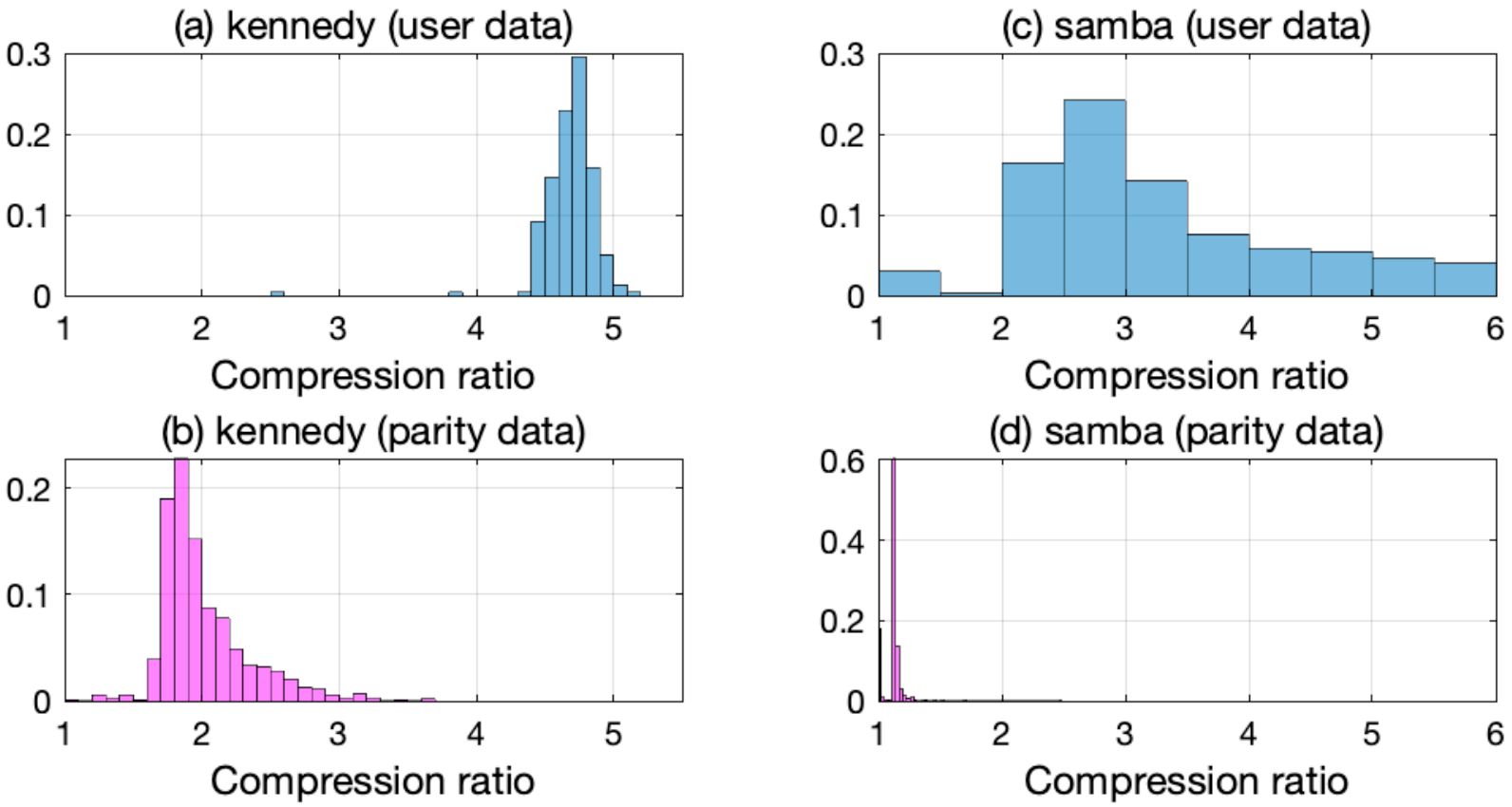}
  \caption{Histograms of per-4KB compression ratio of user data and RAID~5 parity when using two compression benchmark files in the Canterbury corpus~\cite{canterbury-link} and Silesia corpus~\cite{silesia-link}.}
  \label{fig:RAIDratio}
\end{figure}

This subsection presents a technique to further improve the implementation efficiency of elastic RAID. It is motivated by an observation that user data and their RAID parity could have significantly different compressibility. For the purpose of demonstration, using two compression benchmark corpus files (i.e., file {\it kennedy} in the Canterbury corpus~\cite{canterbury-link} and file {\it samba} in the Silesia corpus~\cite{silesia-link}) as user data, we measured the per-4KB compression ratio of both user data and RAID~5 parity (assuming 3$+$1 RAID~5). Fig.~\ref{fig:RAIDratio} shows the histogram of measured per-4KB compression ratio (using the GZIP compression library), which clearly shows that RAID~5 parity has much worse compressibility than user data. For {\it samba}-based experiments, the RAID~5 parity data are even almost completely incompressible. This phenomenon can be explained as follows: Lossless data compression is mainly realized by deduplicating repeated byte strings in the data stream. Different user data strips within the same stripe most likely have repeated byte strings at different locations inside 4KB blocks. As a result, being obtained by bit-wise XOR over multiple user data strips, RAID parity tends to have much less amount of repeated byte strings, leading to a much worse compressibility as illustrated in Fig.~\ref{fig:RAIDratio}. 

This observation suggests that RAID~5 does not necessarily always consume less amount of post-compression physical storage space than RAID~10. Given a RAID~5 stripe over $n+1$ drives, let $\alpha_{usr}$ and $\alpha_{pty}$ denote the compression ratio of user data and RAID parity. The RAID~5 stripe consumes a post-compression physical storage capacity that is proportional to $\frac{n}{\alpha_{usr}}+\frac{1}{\alpha_{pty}}$. In comparison, its RAID~10 counterpart consumes a post-compression physical storage capacity that is proportional to $\frac{2n}{\alpha_{usr}}$. Apparently, if $\alpha_{usr}>n\cdot \alpha_{pty}$, this RAID~5 stripe will consume a larger amount of post-compression physical storage space than its RAID~10 counterpart. Hence, converting this stripe from RAID~5 to RAID~10 is always desirable. This leads to the {\it autonomous RAID level conversion}: Elastic RAID periodically queries SSDs about the compression ratio of user data and parity (i.e., $\alpha_{usr}$ and $\alpha_{pty}$) of each RAID~5 stripe. For all the RAID~5 stripes in which $\alpha_{usr}$ is larger than $n\cdot \alpha_{pty}$, elastic RAID automatically converts them into RAID~10, independent from the scheduling process discussed above in Section~\ref{sec:scheduling}. Of course, once a RAID~5 stripe has been autonomously converted into RAID~10, subsequent data updates may possibly degrade the user data compressibility so that $\alpha_{usr}$ becomes less than $n\cdot \alpha_{pty}$. To simplify the implementation, elastic RAID does not proactively search for such {\it degraded} RAID~10 stripes. Instead, elastic RAID simply leaves it to the scheduler, e.g., once the degraded RAID~10 stripes have become relatively cold, the scheduler will naturally convert them back to RAID~5 if necessary.

\section{Implementation of Elastic RAID}
\label{sec:implementation}
The above presented elastic RAID design strategy could be integrated into a software RAID solution or a dedicated hardware RAID card. For the purpose of demonstration, by modifying the existing Linux mdraid~\cite{Linux-RAID-link}, we implemented a Linux software elastic RAID~(SW eRAID) that supports the dynamic mixture of RAID~5 and RAID~10. It implements a write journal to mitigate the well-known write hole problem~\cite{patterson1988case} of RAID~5 and meanwhile reduce the write request latency experienced by users. The write journal spans over all the SSDs and is protected by RAID~5 as well. Because of the append-only nature of journal, RAID~5 over the write journal is not subject to the write hole problem. Upon receiving a write request, SW eRAID flushes the data into the write journal and then immediately acknowledges the write completion to the user. Data are migrated from the write journal into their destined stripes asynchronously in the background, which is realized in batches with multiple threads in order to prevent background data migration from becoming the overall RAID speed bottleneck. Since currently available SSDs with built-in transparent compression (i.e., the one from ScaleFlux Inc.~\cite{ScaleFlux-link}) have not yet been able to support per-LBA compression ratio query, this prototype SW eRAID does not incorporate the per-stripe compressibility into its RAID level conversion scheduling and does not implement the autonomous RAID level conversion. Once future SSDs support per-LBA compression ratio query, these features could be easily added into the SW eRAID.

The RAID~5 vs.~RAID~10 speed performance difference is most noticeable when serving (i)  random data write requests or (ii) random read requests under degraded mode when one SSD is offline~(e.g., one SSD suffers a catastrophic failure). This is mainly due to the large difference in terms of the read amplification under these two scenarios, which can be explained as follows. Let $n+1$ denote the total number of SSDs in the RAID system. Being protected by RAID~5, the write journal contributes a write amplification of $(n+1)/n$. To update a data block in one RAID~5 stripe, we need to read its old version and the corresponding parity block from SSDs, based on which we calculate the new parity block and then write the new version of both data block and parity block back to SSDs. This leads to a write amplification of 2 and read amplification of 2. In the case of RAID~10, we directly write the two copies of the data block to SSDs, leading to a write amplification of 2 without any read operations. Under the degraded mode when one SSD is offline, RAID~5 typically only serves read requests, while RAID~10 could still serve both read and write requests. In the case of RAID~5, given the total $n+1$ SSDs, $1/(n+1)$ amount of user data are lost and hence must be reconstructed via RAID~5 calculation at the read amplification of $n$. Therefore, the overall read amplification is $2n/(n+1)$. In the case of RAID~10, no user data are lost even if one drive is offline, hence the read amplification is 1. Table~\ref{tab:amplification} summarizes the write/read amplification of the SW eRAID when serving random writes under the normal mode and random reads under the degraded mode. It clearly shows the significant read amplification difference between RAID~5 and RAID~10, which directly results in significant speed performance difference. 

\begin{table}[htbp]
\centering
\caption{Write/read amplification under random data access.}
\label{tab:amplification} 
\begin{tabular}{|c|c|c|c|}
\hline
 & \multicolumn{ 2}{c|}{Random write} & Random read  \\
 & \multicolumn{ 2}{c|}{(normal mode)} & (degraded mode) \\
 \cline{2-4}
 & WA &\hspace*{3pt} RA \hspace*{3pt}& RA \\
 \hline
 RAID~5 & $2+(n+1)/n$ & 2 & $2n/(n+1)$\\
 \hline
 RAID~10 & $2+(n+1)/n$ & 0 & 1\\
\hline
\end{tabular}
\end{table}
\vspace*{-12pt}WA: write amplification; RA: read amplification.

\begin{figure*}[hbtp]
  \centering
  \includegraphics[width = 0.95\linewidth]{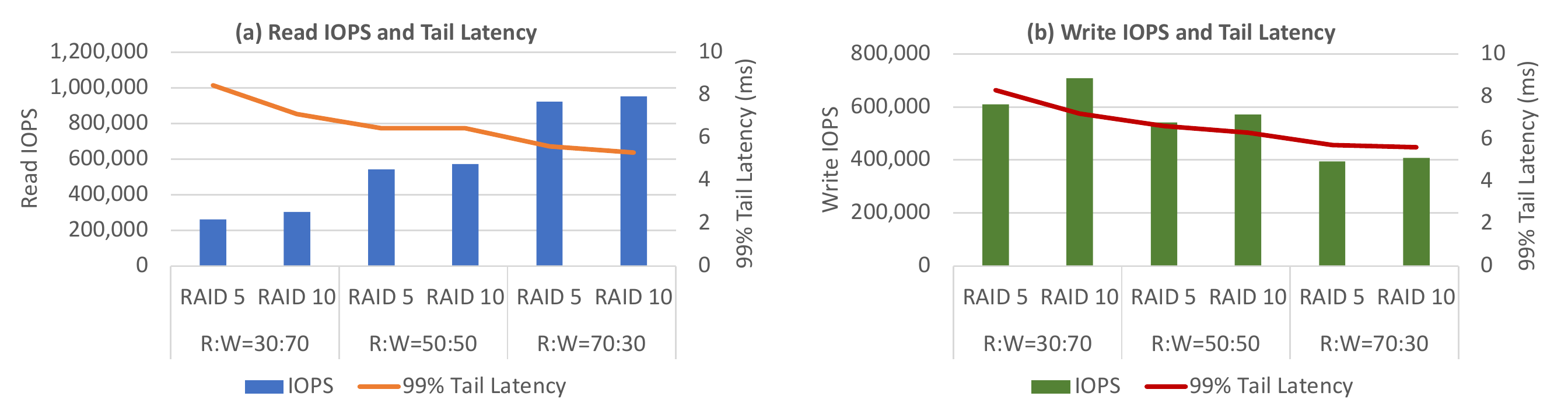}
  \caption{IOPS and 99\% tail latency of (a) read requests and (b) write requests under different FIO 4KB random read/write-mixed workloads.}
  \label{fig:mixed}
\end{figure*}
\section{Evaluation}
\label{sec:evaluation}
To evaluate the effectiveness of the proposed elastic RAID design strategy and study the involved trade-offs, we carried out experiments on a server with a 26-core 2.5GHz Intel Xeon CPU, 565GB DRAM, and commercial NVMe SSDs with built-in transparent compression from ScaleFlux Inc.~\cite{ScaleFlux-link}. The NVMe SSDs carry out hardware-based zlib~\cite{zlib-link} compression on each 4KB LBA data block directly along the I/O path. We applied the widely used FIO~(flexible I/O tester) tool~\cite{FIO-link} to generate heavy foreground random data access I/O workloads and collect the IOPS and tail latency results. The system OS is CentOS Linux release 7.6.1810, and the FIO version is 3.13. In all the experiments, RAID strip size is set as 4KB, and the RAID system contains 4 SSDs.

\subsection{Baseline RAID~5 Performance}
\label{sec:baseline}
We first carried out experiments to evaluate the baseline speed performance under random write workloads when the system operates in the RAID~5 only mode (i.e., all the data are protected by RAID~5). For the purpose of comparison, we also tested the existing Linux mdraid and the state-of-the-art software RAID product RAIDIX~\cite{RAIDIX-link}. We focused on 4KB random write workloads over the entire LBA span to trigger the worst-case scenarios for RAID~5. To ensure sufficient I/O workload stress over all the three RAID~5 systems, we configured FIO to run with 16 jobs at the I/O queue depth of 128. The RAID system contains total 4 SSDs (i.e., 3+1 RAID~5). Table~\ref{tab:baseline} lists the IOPS and tail latency results of the three different RAID~5 systems.
\begin{table}[htbp]
\centering
\caption{RAID~5 performance under 4KB random writes.}
\label{tab:baseline} 
\begin{tabular}{|c|r|r|r|}
\hline
 &\multicolumn{1}{c|}{\multirow{2}*{IOPS}} & \multicolumn{ 2}{c|}{Tail latency (ms)}\\
 \cline{3-4}
 & & \multicolumn{1}{c|}{99\%} 
 &  \multicolumn{1}{c|}{99.9\%}\\
 \hline
 \hspace*{6pt}Linux mdraid \hspace*{6pt} & 59,003 \hspace*{2pt}& \hspace*{6pt}40.1 \hspace*{2pt}&\hspace*{2pt} 43.8 \hspace*{2pt}\\
 \hline
 RAIDIX & \hspace*{6pt} 474,389 \hspace*{2pt}& 40.1 \hspace*{2pt}& \hspace*{6pt} 246.4 \hspace*{2pt}\\
 \hline
 SW eRAID & \hspace*{4pt} 677,829 \hspace*{2pt} & 9.6 \hspace*{2pt}& 14.9 \hspace*{2pt}\\
 \hline
\end{tabular}
\end{table}

The results show that RAIDIX and our SW eRAID can achieve one order of magnitude higher 4KB random write IOPS than Linux mdraid. This is mainly because Linux mdraid uses a single management thread to control the RAID~5 state machine. Although such single-thread management implementation works well on HDD-based RAID, it could easily become the speed performance bottleneck of SSD-based RAID. Compared with RAIDIX, our SW eRAID achieves 1.4$\times$ higher 4KB random write IOPS at 4.2$\times$ shorter 99\% tail latency. Because we do not have access to the source code of RAIDIX, we conjecture that this is mainly due to the different implementation efficiency of background data movement from the write journal to destined stripes. The results well demonstrate that our SW eRAID matches and even exceeds the state of the art of SW RAID implementation.  

\subsection{RAID~5 vs.~RAID~10 Performance}
\label{sec:5vs10performance}
Using our SW eRAID prototype, we evaluated and compared the speed performance when the system operates in RAID~5 or RAID~10 mode.  Table~\ref{tab:eRAIDread} and Table~\ref{tab:eRAIDwrite} list the IOPS and tail latency under 4KB random read and write workloads, respectively. To ensure sufficient I/O workload stress on SSDs, we set 16 FIO jobs at the I/O queue depth of 128. Under random read workloads, regardless RAID~5 or RAID~10, the SSD array serves the same amount of read requests, leading to the similar IOPS and tail latency performance as shown in Table~\ref{tab:eRAIDread}. The slight performance advantage of RAID~10 over RAID~5 is due to the higher flexibility of RAID~10 on choosing which SSD serves each read request. 

\begin{table}[htbp]
\centering
\caption{SW eRAID performance under 4KB random read.}
\label{tab:eRAIDread} 
\begin{tabular}{|c|c|c|c|}
\hline
 \multirow{2}*{RAID Mode} &\multicolumn{1}{c|}{\multirow{2}*{IOPS}} & \multicolumn{ 2}{c|}{Tail latency (ms)}\\
 \cline{3-4}
 & & 99\%  & 99.9\%\\
 \hline
  RAID~5 & \hspace*{6pt}2,469,538 \hspace*{6pt}&  \hspace*{8pt} 1.1 \hspace*{8pt}& \hspace*{8pt} 1.5 \hspace*{8pt}\\
 \hline
  \hspace*{12pt}RAID~10 \hspace*{12pt} & 2,592,096 & 1.0 & 1.4 \\
 \hline
\end{tabular}
\end{table}

In contrast to the case of random read workloads, as shown in Table~\ref{tab:eRAIDwrite}, RAID~10 achieves 1.67$\times$ higher 4KB random write IOPS than RAID~5, while maintaining a similar tail latency. This is due to the significantly less read amplification of RAID~10 compared with RAID~5 under random write workloads as discussed above in Section~\ref{sec:implementation}.
\begin{table}[htbp]
\centering
\caption{SW eRAID performance under 4KB random write.}
\label{tab:eRAIDwrite} 
\begin{tabular}{|c|r|r|r|}
\hline
 \multirow{2}*{RAID Mode} &\multicolumn{1}{c|}{\multirow{2}*{IOPS}} & \multicolumn{ 2}{c|}{Tail latency (ms)}\\
 \cline{3-4}
 & & \multicolumn{1}{c|}{99\%} &  \multicolumn{1}{c|}{99.9\%} \\
 \hline
 RAID~5 & \hspace*{12pt} 677,829 \hspace*{4pt} & \hspace*{8pt} 9.6 \hspace*{4pt}& \hspace*{8pt} 14.9 \hspace*{4pt}\\
 \hline
 \hspace*{12pt} RAID~10 \hspace*{12pt} & 1,138,546 \hspace*{4pt}& 10.0 \hspace*{4pt}& 16.5 \hspace*{4pt}\\
 \hline
\end{tabular}
\end{table}

Fig.~\ref{fig:mixed} shows the measured read/write IOPS and 99\% tail latency under FIO 4KB random read/write-mixed workloads with three different R:W ratios including 30:70, 50:50, and 70:30. The results show that the RAID~10 over RAID~5 performance advantage increases as the I/O workloads become more write-intensive. For example, under the R:W ratio of 30:70, the read/write IOPS of RAID~10 is 1.16$\times$ higher than that of RAID~5, and the difference drops to 1.03 under the R:W ratio of 70:30. This is because RAID~10 vs.~RAID~5 speed performance difference mainly comes from the read amplification difference under random writes.

The above results were obtained under the normal conditions where all the SSDs are functional. Once one SSD becomes off-line (e.g., due to a catastrophic hardware failure), the RAID system will operate in the {\it degraded} mode and serve read requests only. Given the total 4 SSDs in the system, as discussed above in Section~\ref{sec:implementation}, RAID~5 suffers from a read amplification of 1.6 under the degraded mode, while RAID~10 does not experience any read amplification. Therefore, under the degraded mode, RAID~10 should be able to achieve much higher random read speed performance than RAID~5. Table~\ref{tab:degraderead} lists the measured 4KB random read IOPS and tail latency when RAID~5 and RAID~10 operate in the degraded mode. To further stress the SSDs under read workloads, we increase the FIO job number from 16 to 32 while maintaining the same queue depth of 128. The results well demonstrate the advantage of RAID~10 over RAID~5 when serving random read requests in the degraded mode. 

\begin{table}[htbp]
\centering
\caption{4KB random read performance in degraded mode.}
\label{tab:degraderead} 
\begin{tabular}{|c|c|r|r|}
\hline
 \multirow{2}*{RAID Mode} &\multicolumn{1}{c|}{\multirow{2}*{IOPS}} & \multicolumn{ 2}{c|}{Tail latency (ms)}\\
 \cline{3-4}
 & & \multicolumn{1}{c|}{99\%} & \multicolumn{1}{c|}{99.9\%} \\
 \hline
  RAID~5 & \hspace*{6pt} 1,434,445 \hspace*{6pt}&  \hspace*{12pt} 14.5 \hspace*{4pt}&  \hspace*{8pt} 22.9 \hspace*{4pt}\\
 \hline
  RAID~10 & 2,505,795 & 4.4 \hspace*{4pt}& 4.8 \hspace*{4pt}\\
 \hline
\end{tabular}
\end{table}

\subsection{SW eRAID with mixed RAID~5/10}
\label{sec:mixtureperformance}
\begin{figure}[b]
  \centering
  \includegraphics[width = \linewidth]{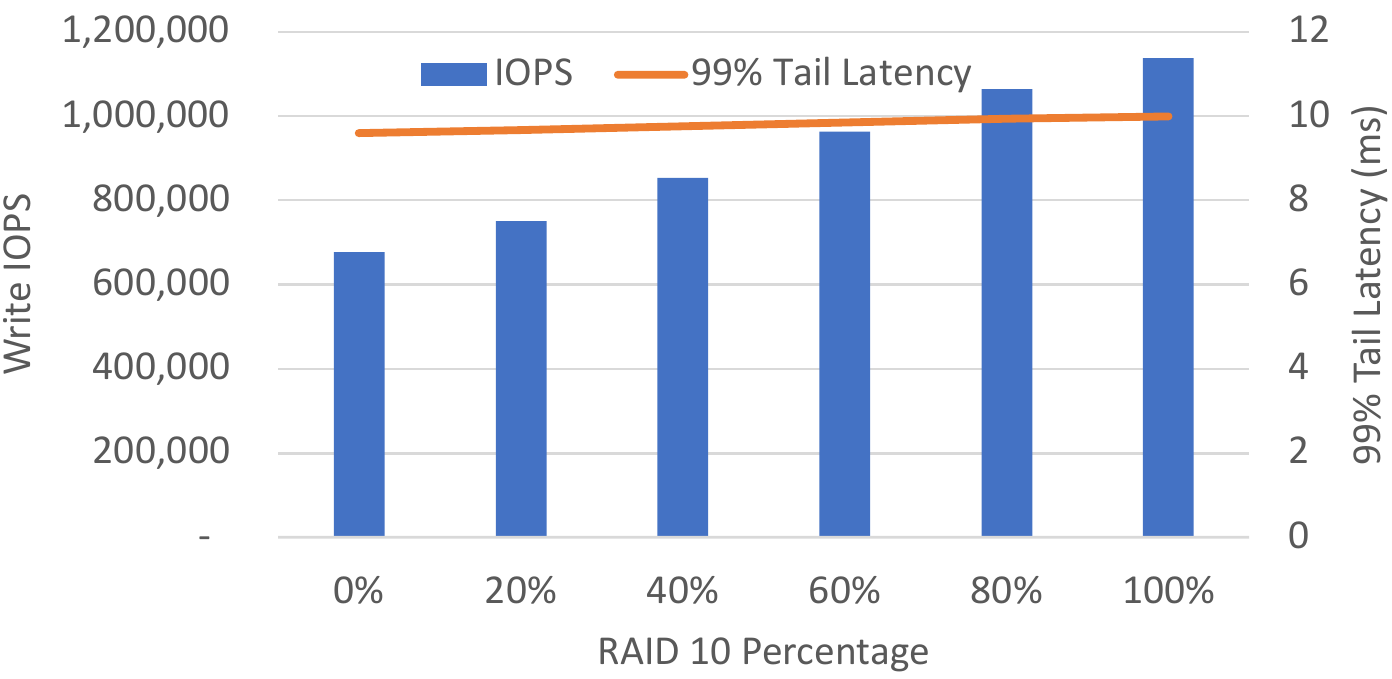}
  \caption{SW eRAID 4KB random write IOPS and 99\% tail latency under different percentage of data being protected by RAID~10.}
  \label{fig:writemixed}
\end{figure}
Given sufficient runtime user data compressibility, SW eRAID could transparently convert data protection from RAID~5 to RAID~10 without sacrificing the effective RAID storage capacity. As discussed and demonstrated above, RAID~10 outperforms RAID~5 mainly in the presence of random write workloads and degraded mode. Fig.~\ref{fig:writemixed} shows the 4KB random write IOPS and tail latency when different percentages of data are protected by RAID~10. As user data compressibility improves, eRAID will convert the protection of more data from RAID~5 to RAID~10, which will lead to a higher overall RAID system I/O speed performance as shown in Fig.~\ref{fig:writemixed}.

\begin{figure}[hbtp]
  \centering
  \includegraphics[width = \linewidth]{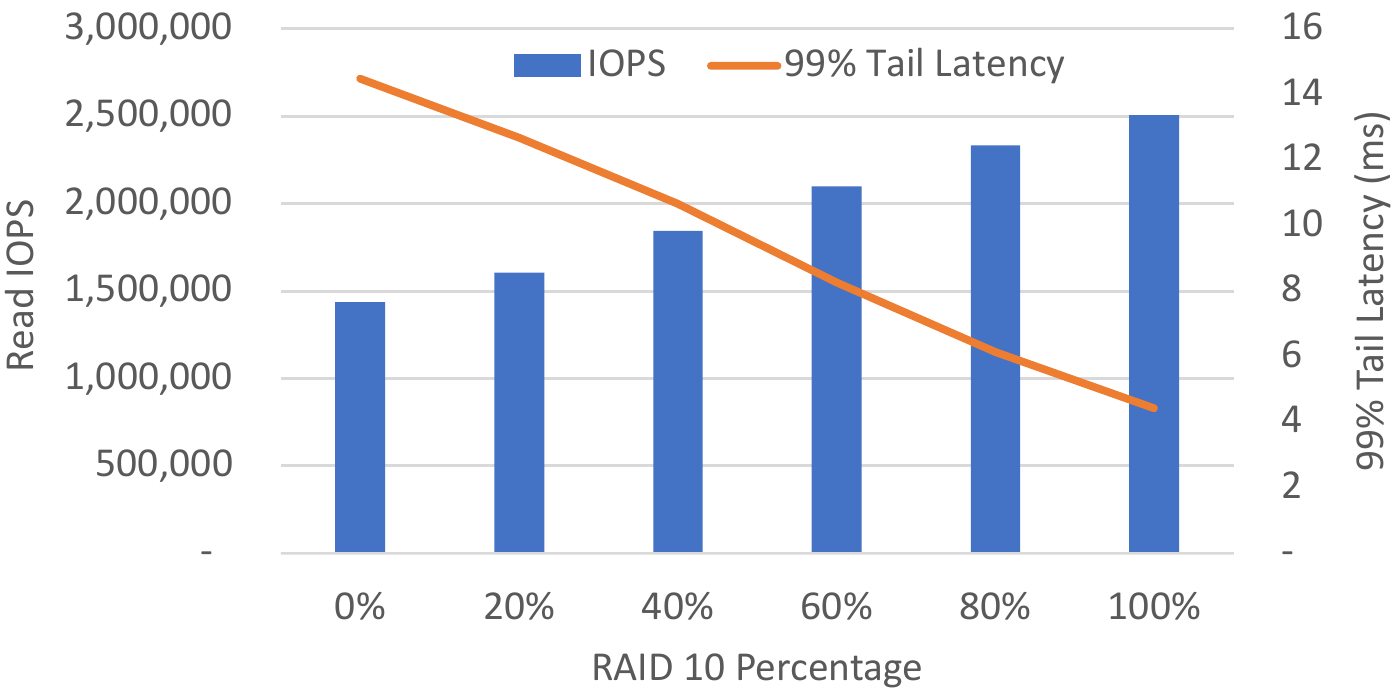}
  \caption{SW eRAID 4KB random read IOPS and 99\% tail latency in the degraded mode under different percentage of data being protected by RAID~10.}
  \label{fig:readmixed}
\end{figure}

Fig.~\ref{fig:readmixed} shows the 4KB random read IOPS and tail latency under the degraded mode when different percentages of data are protected by RAID~10. As one could intuitively justify, the read speed performance will improve when more data are protected under RAID~10 under a higher user data compressibility. As shown in Fig.~\ref{fig:readmixed}, the 99\% read tail latency largely drops even as the IOPS increases. This is because a higher percentage of RAID~10 leads to a lower read amplification and hence a smaller read I/O queue depth, which directly contributes to shorter read latency. 

\begin{figure}[t]
  \centering
  \includegraphics[width = \linewidth]{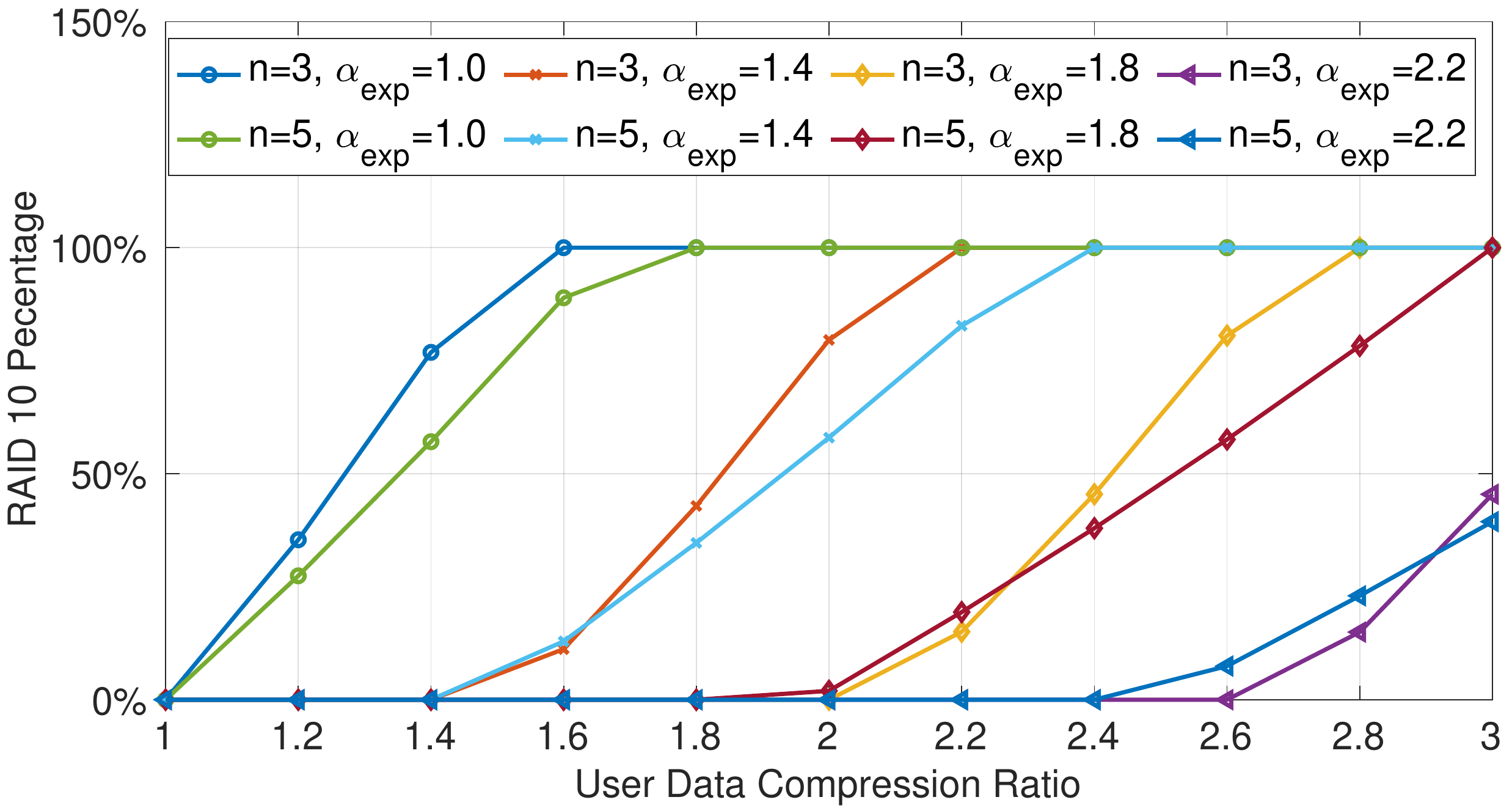}
  \caption{Percentage of RAID~10 under different storage capacity expansion factor $\alpha_{exp}$.}
  \label{fig:RAID10percnt1}
\end{figure}

The above results show that the performance advantage of SW eRAID heavily depends on the percentage (denoted as $P_{conv}$) of data being converted from RAID~5 to RAID~10. In addition to the user data compressibility, $P_{conv}$ also depends on the storage capacity expansion factor $\alpha_{exp}\ge 1$ and storage capacity utilization factor $\beta_{util}\le 1$. Given the total $n+1$ SSDs and per-SSD physical storage capacity of $C_{flash}$, eRAID exposes a total logical storage capacity of $\alpha_{exp}\cdot n\cdot C_{flash}$ to the user. As we increase $\alpha_{exp}$ to more aggressively leverage data compressibility to increase the effective RAID storage capacity, less amount of residual data compressibility will be left for RAID~5 to RAID~10 conversion. Meanwhile, as discussed above in Section~\ref{sec:scheduling}, users may not always utilize 100\% of the storage capacity, and eRAID could accordingly schedule the RAID level conversion in response to the runtime storage capacity utilization factor $\beta_{util}$.  Fig.~\ref{fig:RAID10percnt1} shows the percentage of RAID~10 under different storage capacity expansion factor $\alpha_{exp}$. We considered the cases of $n=3$ or $n=5$ (i.e., the RAID systems contains 4 or 6 SSDs). The storage capacity utilization factor $\beta_{util}$ is set as 100\% (i.e., user fully utilizes the entire storage capacity exposed by the RAID system). As discussed above in Section~\ref{sec:optimization}, RAID~5 parity tends to have much worse compression ratio than user data. Let $\alpha_{usr}$ and $\alpha_{pty}$ denote the compression ratio of user data and RAID~5 parity, according to the results presented in Section~\ref{sec:optimization}, we set $(\alpha_{usr}-1)=4(\alpha_{pty}-1)$ so that the RAID~5 parity is incompressible (i.e., $\alpha_{pty}=1$) when user data is incompressible (i.e., $\alpha_{usr}=1$), and RAID~5 parity compression ratio is 1.5 when user data compression ratio is 3. As shown in Fig.~\ref{fig:RAID10percnt1}, 
as we increase the storage capacity expansion factor $\alpha_{exp}$, a larger user data compression ratio is required to enable RAID~5 to RAID~10 conversion. Under a larger storage capacity expansion factor $\alpha_{exp}$, the RAID~10 percentage will increase more slowly with the user data compression ratio $\alpha_{usr}$. For example, when $\alpha_{exp}=1.0$, the user data compression ratio should increase from 1 to 1.6 in order to enable the 100\% RAID~5 to RAID~10 conversion; when $\alpha_{exp}=1.8$, the user data compression ratio must jump from 1.8 to 2.8 in order to enable 100\% RAID~5 to RAID~10 conversion. The results also show the noticeable impact of the total number of SSDs on the relationship between user data compressibility and RAID~10 coverage percentage, i.e., the more SSDs are deployed in the RAID system, the slower the RAID~10 coverage percentage will grow with the user data compression ratio.

\begin{figure}[htbp]
  \centering
  \includegraphics[width = \linewidth]{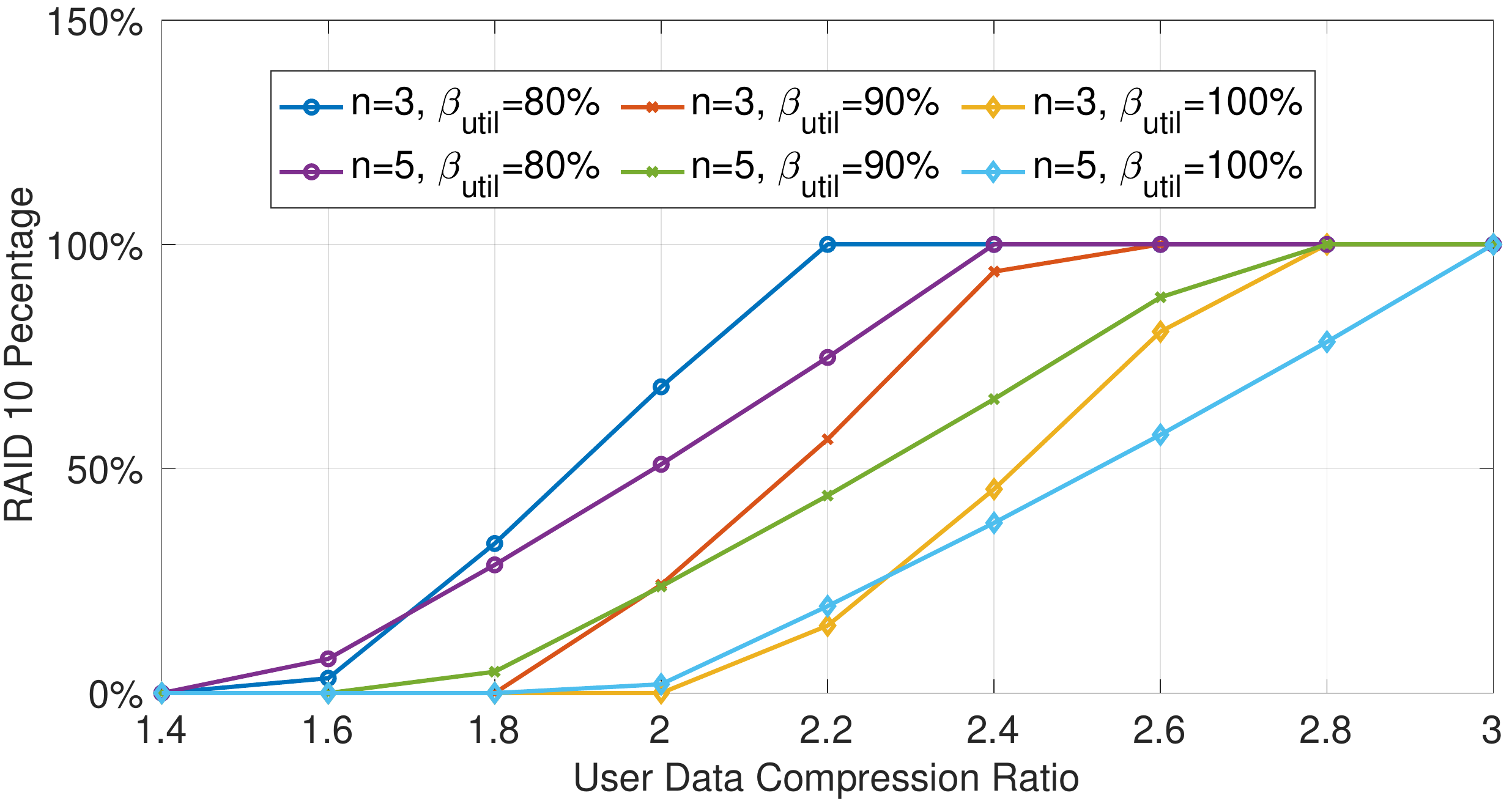}
  \caption{Percentage of RAID~10 under different storage capacity utilization factor $\beta_{util}$.}
  \label{fig:RAID10percnt2}
\end{figure}

Fig.~\ref{fig:RAID10percnt2} shows the percentage of RAID~10 under different storage utilization factor $\beta_{util}$. We set the storage capacity expansion factor $\alpha_{exp}$ as 1.8. The results show that under-utilized storage capacity could be effectively leveraged to improve the RAID~10 coverage percentage. For example, under the user data compression ratio of 2, the RAID~10 coverage percentage is almost 0\% when the storage capacity utilization factor $\beta_{util}$ is 100\%, and it will improve to 25\% and 68\% when $\beta_{util}$ reduces to 90\% and 80\%, respectively. Since it is not uncommon for real production environment to have a storage capacity utilization factor of 80\%$\sim$90\% or even below, the results suggest that the adaptive RAID level conversion scheduling could very noticeably improve the RAID~10 coverage percentage.

\begin{figure}[hbtp]
  \centering
  \includegraphics[width = \linewidth]{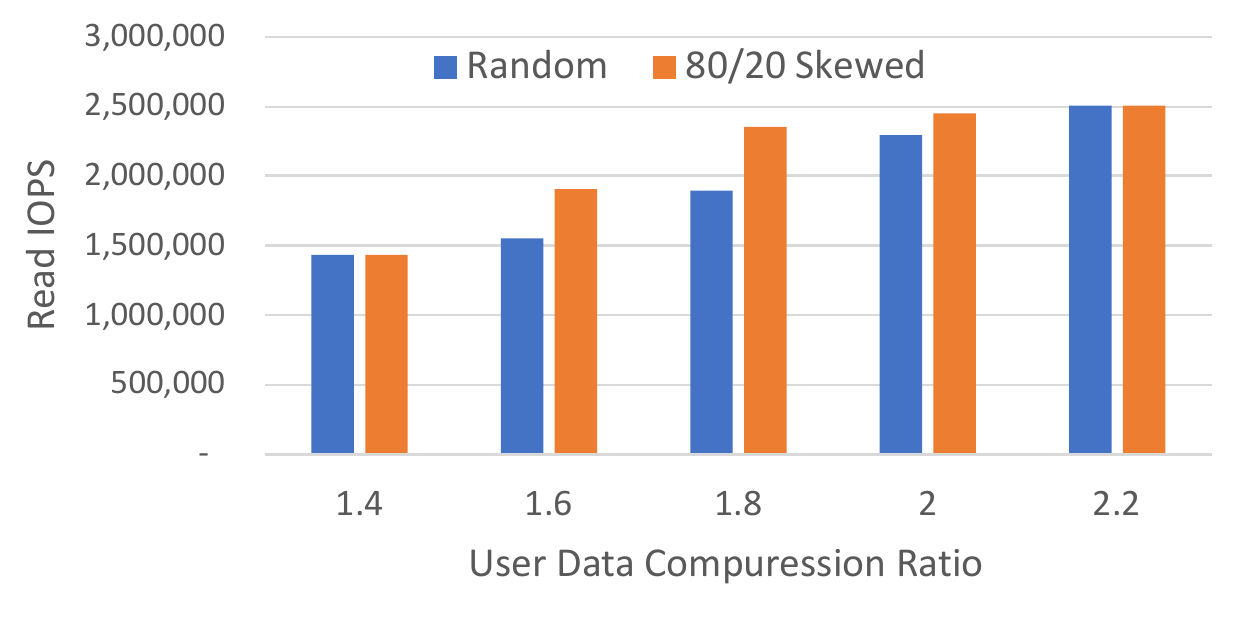}
  \caption{SW eRAID 4KB random read IOPS in the degraded mode under random and 80/20 skewed workloads.}
  \label{fig:skewed}
\end{figure}

The above results show that a small increase of user data compression ratio could enable noticeable increase of RAID~10 coverage percentage. Meanwhile, real-world I/O workloads typically exhibit considerable degrees of access locality. As discussed above in Section~\ref{sec:scheduling}, eRAID could schedule RAID level conversion in adaptation to the runtime data access locality. Fig.~\ref{fig:skewed} shows the 4KB random read IOPS in the degraded mode under random workload (i.e., zero data access locality) and 80/20 skewed workload (i.e., 80\% of reads hit 20\% of data). The eRAID system storage capacity expansion factor $\alpha_{exp}$ is set to 1.4, and the RAID storage capacity utilization factor $\beta_{util}$ is set to 100\%. The RAID~10 coverage percentage increases from 0\% to 100\% when the user data compression ratio increases from 1.4 to 2.2. As shown in Fig.~\ref{fig:skewed}, under the 80/20 skewed workload, read IOPS could improve by 33\% even when the user data compression ratio only increases from 1.4 to 1.6. 

\subsection{RAID Level Conversion}
We further evaluated the RAID level conversion and its impact on RAID system I/O speed performance. As discussed above, our SW eRAID implementation uses multiple background threads to carry out RAID level conversion between RAID~5 and RAID~10. The RAID level conversion throughput improves as we increase the number of RAID level conversion threads. Table~\ref{tab:conversion} shows the RAID level conversion throughput in the absence of foreground user read/write requests. The RAID system contains 4 SSDs, and the user data compression ratio is set as 2:1. As discussed above in Section~\ref{sec:RAIDimplementation}, RAID level conversion is performed on the per-stripe basis. To convert one stripe from RAID~5 to RAID~10, we only need to duplicate the data once within the stripe segment; to convert one stripe from RAID~10 to RAID~5, we need to fetch the data stripe, and calculate and write the RAID~5 parity. Hence, RAID~10 to RAID~5 conversion tends to have a lower throughput than RAID~5 to RADI~10 conversion, as shown in Table~\ref{tab:conversion}. The results show that the RAID level conversion can easily achieve multi-GB/s throughput~(without foreground user read/write requests), which well demonstrates the efficiency of the proposed elastic RAID implementation strategy.

\begin{table}[t]
\centering
\caption{RAID level conversion throughput.}
\label{tab:conversion} 
\begin{tabular}{|c|c|c|}
\hline
Conversion & \# of conversion & Throughput \\
mode & threads & (GB/s)\\
\hline\hline
 \multirow{2}*{RAID~5 to RAID~10} & 8 & 3.1\\
 \cline{2-3}
 & 24 & 6.4\\
 \hline\hline
 \multirow{2}*{RAID~10 to RAID~5} & 8 & 2.6\\
 \cline{2-3}
 & 24 & 5.1\\
 \hline
\end{tabular}
\end{table}

\begin{figure}[b]
    \centering
  \includegraphics[width = \linewidth]{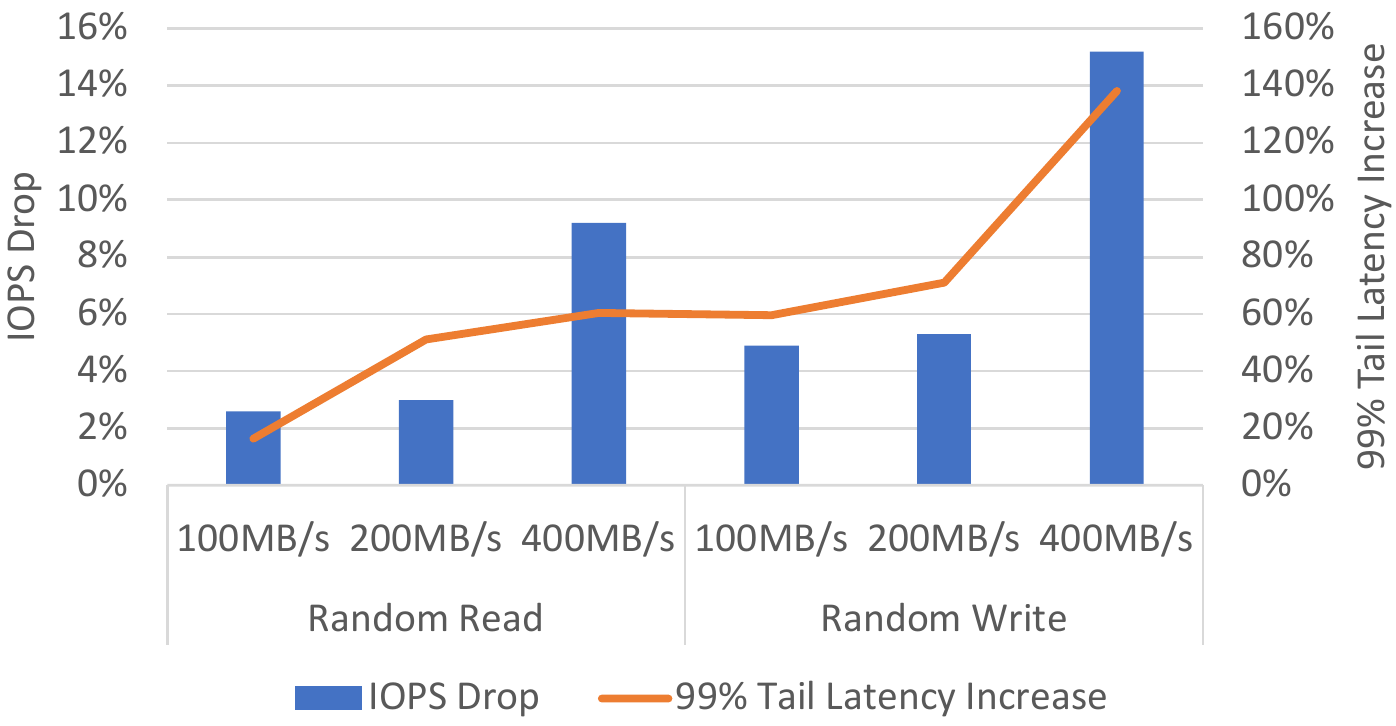}
  \caption{Impact of background RAID~5 to RAID~10 conversion on heavy random read or write workloads with 32 FIO jobs.}
  \label{fig:impact5}
\end{figure}

Next, we evaluated the impact of background RAID level conversion on the performance of SW eRAID serving foreground user read/write requests.  It is reasonable to expect that average user data compression ratio typically does not significantly vary over a short period (e.g., it is unlikely for the average user data compression ratio on a 32TB RAID system drops from 2 to 1.2 within just couple hours). Therefore, we could control the performance impact by throttling the background RAID level conversion throughput without compromising the accuracy of RAID~10 coverage. Meanwhile, the impact of background RAID level conversion also depends on the intensity of the foreground user read/write requests, i.e., the heavier the foreground I/O workloads are, the more severely they will be impacted by the background RAID level conversion. To exam the worst-case scenario, we use heavy 4KB random read or write workloads with 32 FIO jobs to emulate very heavy foreground user I/O workloads.
  \begin{figure}[t]
  \centering
  \includegraphics[width = \linewidth]{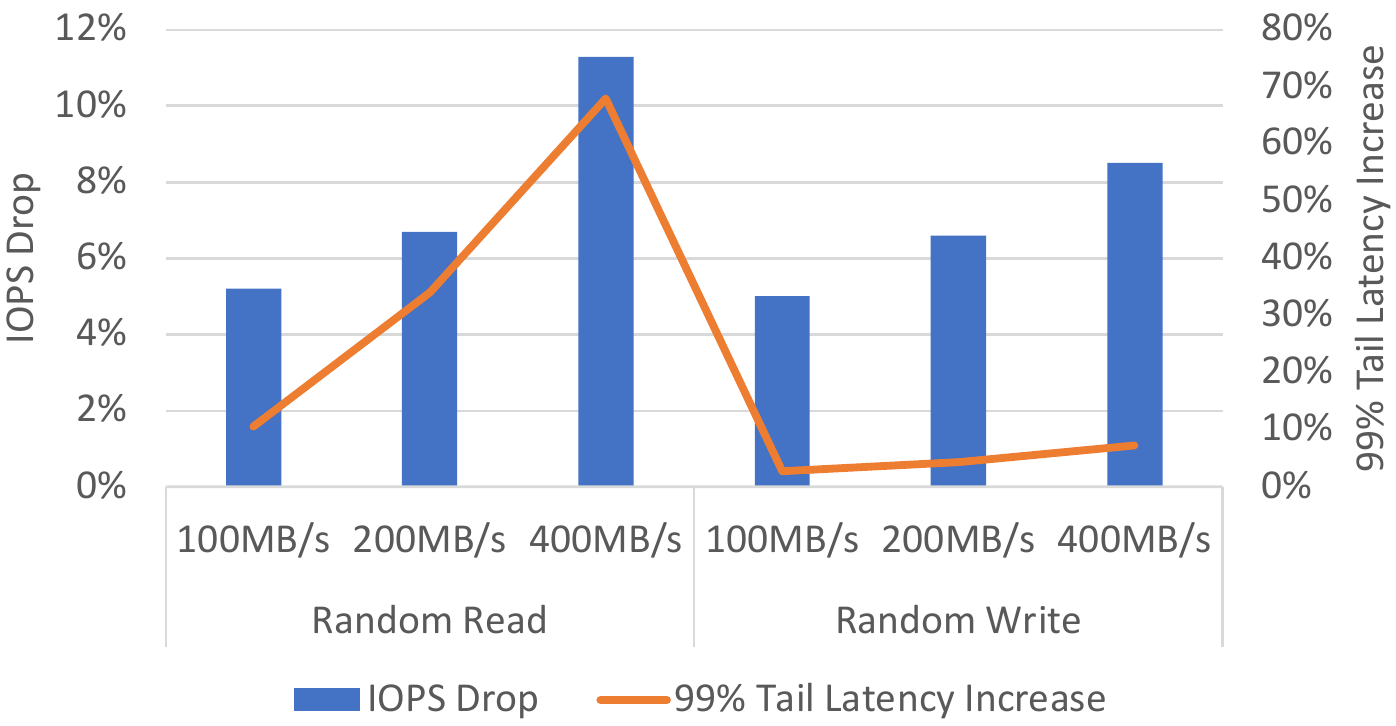}
  \caption{Impact of background RAID~10 to RAID~5 conversion on heavy random read or write workloads with 32 FIO jobs.}
  \label{fig:impact10}
\end{figure}
Fig.~\ref{fig:impact5} and Fig.~\ref{fig:impact10} show the impact of background RAID level conversion under three different throttled conversion throughput, including 100MB/s, 200MB/s, and 400MB/s. The results show that, even under the very heavy 32-job FIO foreground I/O workloads, the impact of background RAID level conversion can be small~(e.g., only few percentage of IOPS drop with 200MB/s background conversion). The results show that, compared with RAID~5 to RAID~10 conversion, RAID~10 to RAID~5 conversion causes similar impact on random read speed performance but causes noticeably less impact on random write speed performance. This is because, compared with RAID~5 to RAID~10 conversion, RAID~10 to RAID~5 conversion generates much less amount write traffic to SSDs.

\subsection{Impact of Stripe Selection Accuracy}
\begin{figure}[hbtp]
  \centering
  \includegraphics[width = \linewidth]{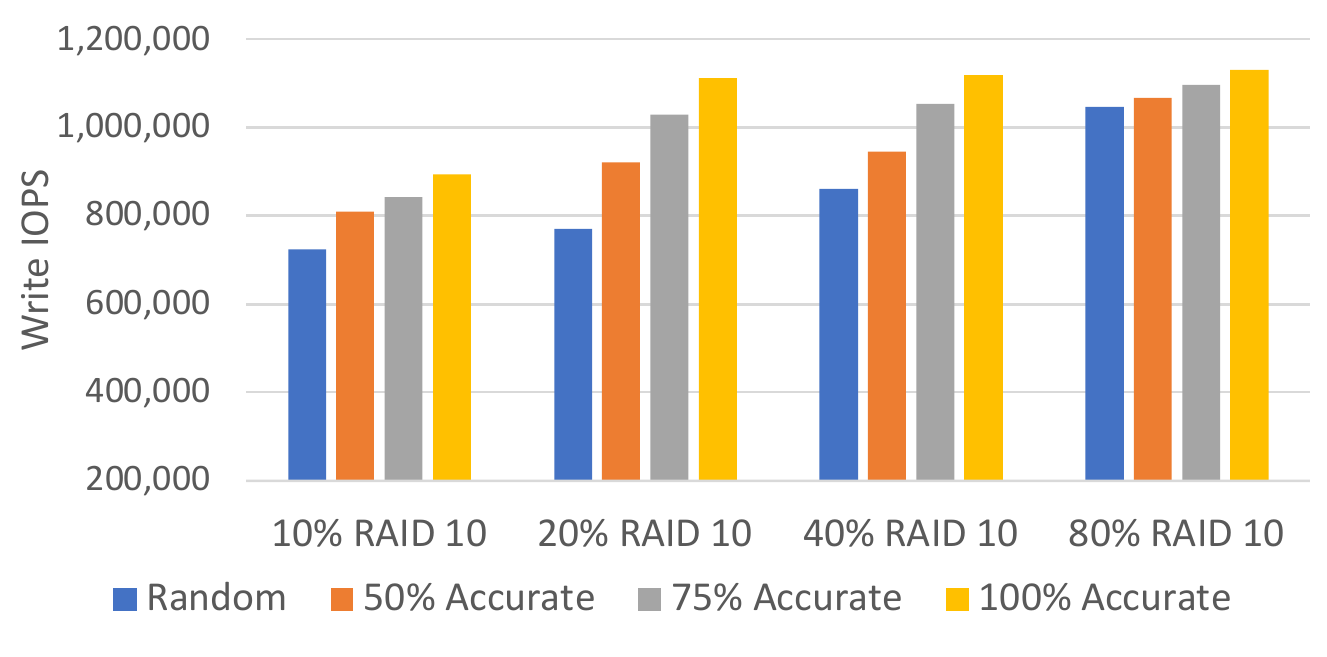}
  \caption{Impact of data classification accuracy on 4KB random write IOPS of 80/20 skewed workloads.}
  \label{fig:impactconversion}
\end{figure}

Finally, we studied the impact of non-ideal stripe selection for RAID level conversion. As discussed above in Section~\ref{sec:scheduling}, elastic RAID should keep read/write-hot stripes in RAID~10, for which a data access classification algorithm (e.g., second chance or LRU) must be used. Our SW eRAID prototype chose the simple and low-cost second chance algorithm. Regardless which classification algorithm is chosen, its accuracy strongly depends on the runtime I/O workload characteristics, i.e., the more quickly the I/O data access locality varies, the less accurate the runtime data classification will be. We carried out experiments to evaluate the impact of data classification accuracy. We use two FIO processes to emulate 80/20 skewed 4KB random write workloads. Fig.~\ref{fig:impactconversion} shows the overall random write IOPS under different data classification accuracy, e.g., 75\% {\it accuracy} means the data classification algorithm captures 75\% of write-hot data under the 80/20 skewed 4KB random write workloads. For the purpose of comparison, we also considered a {\it Random} case where data are just randomly classified as write-hot. We studied four different data compressibility under which 10\%, 20\%, 40\%, and 80\% of data can be converted into RAID~10. As the data classification accuracy improves, a higher percentage of write-hot data will be covered by RAID~10, leading to a higher write IOPS as shown in Fig.~\ref{fig:impactconversion}. Moreover, as data compressibility increases to enable a higher RAID~10 coverage, data classification accuracy tends to have a less impact on the overall IOPS. For example, when 20\% of data can be converted into RAID~10, write IOPS could increase by 20.6\% if the data classification accuracy improves from 50\% to 100\%. In comparison, when 80\% of data can be converted into RAID~10, increasing the data classification accuracy from 50\% to 100\% improves write IOPS by only 6\%. This further justifies the use of low-cost data classification algorithms (e.g., the second chance algorithm) in the practical implementation.

\section{Related Work}
\label{sec:related}
\noindent{\bf RAID}. As the most widely deployed data protection solution, RAID has been very well studied in the open literature. Prior work on RAID mainly focused on more accurately modeling and analyzing the RAID system reliability~\cite{baek2001reliability, elerath2008highly, thomasian2009higher, li2013stochastic, elerath2014beyond, machida2015performability}, reducing the drive rebuild time and accelerating the data recovery process~\cite{Muntz-RAID-analysis90, holland1992parity, wan2010s2, xiang2010optimal, wang2014mdr}, and better embracing the device characteristics of SSDs~\cite{balakrishnan2010differential, im2010flash, moon2013don, chan2018elastic}. Prior research~\cite{wilkes1996hp, RAID-tiering-Muppalaneni-00} also studied the implementation of {\it tiered} RAID systems consisting of a hot-data RAID~10 tier and warm/cold-data RAID~5/6 tier. The well-known AutoRAID~\cite{wilkes1996hp} has been commercialized by HP from late 1990s to late 2000s.  Operating on traditional HDDs or SSDs, tiered RAID systems must employ complicated data management/migration strategies to realize dynamic tiering and embrace different data mapping functions of different RAID levels, and they are still fundamentally subject to the speed performance vs.~storage cost trade-off. These are at least part of the reason why HP terminated the AutoRAID product line. To our best knowledge, no prior work has ever studied the feasibility of opportunistically leveraging runtime data compressibility to enable dynamic mixture of different RAID levels without sacrificing the effective data storage cost.\\[6pt]
\noindent{\bf Transparent Data Reduction}. Prior research has well studied the implementation of storage data reduction (e.g., compression and deduplication) with complete transparency to user applications. Transparent data reduction can be realized at the filesystem level~\cite{burrows1992line, bonwick2003zettabyte, srinivasan2012idedup, rodeh2013btrfs}, block level~\cite{klonatos2012transparent, VDO-link, ajdari2019cidr}, and even inside storage hardware~\cite{chen2011caftl, wu2012delta, zuck2014compression, chen2019simultaneously}. 
Modern all-flash array products~(e.g., 
Pure Storage FlashBlade~\cite{Pure-Storage-link}) always come with the built-in hardware-based transparent compression capability. Cloud vendors have started to integrate hardware-based compression capability into their storage infrastructure, e.g., Microsoft Corsia~\cite{Microsoft-acceleration-19} and emerging DPU (data processing unit)~\cite{Nvidia-DPU-HotChip-21}.
Motivated by the emergence of storage hardware with built-in transparent compression, researchers have recently studied its implications to the design of hash-based key-value store~\cite{chen2021kallaxdb} and B-tree~\cite{yifan2022fast}.\\[6pt]
\noindent{\bf SSD Sparse Addressing}. Prior work studied how one could innovate the data management systems by making SSD expose a sparse logical address space that is (much) larger than its internal physical storage capacity.  FlashTier~\cite{saxena2012flashtier} utilizes the SSD sparse addressing to largely simplify the SSD cache management. FlashMap~\cite{huang2015unified} integrates virtual address translation and SSD FTL sparse address translation to efficiently support memory-mapped SSD files. Das et al.~\cite{das2014nvm} presented a solution that leverages the SSD sparse addressing to facilitate application-level data compression.  DFS filesystem~\cite{josephson2010dfs} takes advantage of SSD sparse addressing to significantly simplify its data management.

\section{Conclusions}
\label{sec:conclusions}
This paper presents an elastic RAID design strategy to take full advantage of modern SSDs with the built-in transparent compression capability. The key idea is to leverage the opportunistically available residual data compressibility to enable dynamic RAID~5 to RAID~10 conversion, which can improve the RAID system speed performance without sacrificing its effective storage capacity. This paper develops a bloated stripe allocation design technique to enable simple and fine-grained RAID level conversion by exploring the LBA space expansion capability of SSDs with built-in transparent compression, and presents a dynamic RAID level scheduling strategy that can help to realize the full potential of elastic RAID on improving the speed performance. This paper also presents a simple autonomous RAID level conversion method to further improve the effectiveness of elastic RAID.  We implemented a software-based elastic RAID prototype in support of dynamic conversion between RAID~5 and RAID~10, and experimental results well demonstrate the effectiveness of the proposed elastic RAID design solution. 

\bibliographystyle{abbrv}
\bibliography{reference}
\end{document}